\documentstyle[11pt,newpasp,twoside,psfig]{article}
\def\edcomment#1{\iffalse\marginpar{\raggedright\sl#1\/}\else\relax\fi}
\reversemarginpar

\def\farcs{\hbox{$.\!\!^{\prime\prime}$}}
\begin{document}
\title{Two-dimensional kinematics and stellar populations of early-type 
galaxies: First results from the {\tt SAURON} survey}
\author{Roland Bacon$^1$, Martin Bureau$^6$, Michele Cappellari$^5$, 
Yannick Copin$^7$, Roger Davies$^2$, Eric Emsellem$^1$, Harald Kuntschner$^3$, 
Richard McDermid$^2$, Bryan Miller$^8$, Reynier Peletier$^{1,4}$, Ellen 
Verolme$^5$ \& Tim de Zeeuw$^5$}

\affil{$^1$ Centre de Recherche Astronomique de Lyon, Saint-Genis-Laval, France}
\affil{$^2$ Physics Department, University of Durham, Durham, UK}
\affil{$^3$ European Southern Observatory, Garching bei M\"unchen, Germany}
\affil{$^4$ Department of Physics and Astronomy, University of Nottingham, 
Nottingham, UK}
\affil{$^5$ Sterrewacht Leiden, Leiden, The Netherlands}
\affil{$^6$ Department of Astronomy, Columbia University, New York, USA}
\affil{$^7$ Institut de Physique Nucl\'eaire de Lyon, Villeurbanne, France}
\affil{$^8$ Gemini Observatory, La Serena, Chile}

\begin{abstract}
We present the {\tt SAURON} project, which is aimed at studying the
morphology, two-dimensional kinematics and stellar populations of a
representative sample of elliptical galaxies and spiral bulges. {\tt
SAURON}, a dedicated integral-field spectrograph that is optimized for
wide-field observations and has high throughput, was built in Lyon and is
now operated at the WHT 4.2m telescope. At present, we have observed
approximately two thirds of the seventy-two sample galaxies with {\tt
SAURON}. A comparison with published long-slit measurements
demonstrates that the {\tt SAURON}-data is of equal or better quality,
and provides full two-dimensional coverage.  The velocity and velocity
dispersion fields exhibit a large variety of morphologies: from simple
rotating systems to cylindrical, disky and triaxial velocity fields,
bars and decoupled cores. Most of these kinematical signatures do not
have counterparts in the light distribution. While some 
galaxies are consistent with axisymmetry, most are more complex
systems than assumed previously.  This suggests that the kinematical
properties of nearby E/S0 galaxies do not agree with the often assumed
simplistic two-family model, in which the giant non-rotating triaxial
ellipticals are opposed to the fast-rotating axisymmetric faint
ellipticals and S0s. 
%
\end{abstract}

\section{Introduction}

The formation and evolution of galaxies is an important topic of study
for the main next-generation space (e.g., NGST) and ground-based
(e.g., VLT) telescopes. The coming decade will provide a wealth of
high-quality observational data on high redshift galaxies.  However, a
full understanding of galaxy formation and evolution can only be
achieved by completing our knowledge of the present status of
galaxies.

According to the `standard' scenario, elliptical galaxies and spiral
bulges are the result of complex processes, including successive
mergers and star formation events. Evidence for this scenario can be
found by studying the intrinsic shapes and internal dynamics of these
objects. In the past years, a number of photometric and spectroscopic
studies using ground-based telescopes and HST were devoted to
this. From these studies, two classes of spheroidal galaxies emerge:
\begin{itemize}
\vskip -\baselineskip
\vskip 15 truept
\itemsep -3pt
\item {\bf Giant ellipticals}. These galaxies are red, have a high
metal content and boxy isophotes, are supported by anisotropic
velocity distributions and have triaxial figures. Their nuclear
luminosity profiles have shallow cusps.
\item {\bf Low-luminosity ellipticals and spiral bulges}. These
objects are blue, less metal rich, have disky isophotes, are flattened
by rotation and have nearly-oblate shapes. Their central luminosity
profiles show steep cusps.
\end{itemize}
\vskip -3truept
Until recently, kinematical studies were conducted with long-slit
spectrographs. The limited spatial coverage of these instruments is
not sufficient to capture the complicated velocity fields of triaxial
galaxies.  Furthermore, the few two-dimensional kinematical
observations that were obtained with previous integral-field
spectrographs show more complexity than often assumed, even for
objects that were supposedly axisymmetric (e.g. M31, Bacon et al.\
2001 and references therein).

The {\tt SAURON} project is aimed to study the morphology,
two-dimensional kinematics and stellar populations of a representative
sample of early-type galaxies. To achieve this goal we have designed
and built a dedicated panoramic integral-field spectrograph and
developed new software for data reduction, analysis and
state-of-the-art modeling. Here we present the project and describe
some preliminary results, with emphasis on the stellar kinematics.
Other contributions to this volume present different aspects of the
project, including preliminary results on line-strengths.

\section{Instrument}
The instrument is a lenslet integral-field spectrograph, similar to
the {\tt TIGER} (Bacon et al., 1995) and {\tt OASIS} (Bacon et al.,
2000) instruments that have been operated successfully at CFHT.  By
contrast to {\tt OASIS}, which is optimized for high spatial
resolution, {\tt SAURON} was designed to have a large field of view
(Table~1).  The instrument incorporates a system for obtaining
simultaneous sky spectra to ensure accurate sky subtraction. {\tt
SAURON} was commissioned at the William Herschel Telescope in early
February 1999 (Fig.~1).

The current spectral window is 4810--5340~\AA, a wavelength range that
contains absorption lines for stellar dynamical and line-strength
measurements (H$\beta$, Mg and Fe) and emission lines for ionized gas
studies (H$\beta$, [OIII]). The full spectral range of {\tt SAURON}
covers 4500--7200~\AA.

\begin{center}     
\begin{table}
\caption{{\tt SAURON} specifications}
\begin{tabular}{ll}
\hline
Telescope & William Herschel (4.2m) \\[5pt]
Field of view & $33'' \times 41''$  (LR) \\
& $9'' \times 11''$ (HR) \\
Sampling & $0\farcs94$ (LR)\\
& $0\farcs27$ (HR) \\
Lenslets & 1577 (of which 146 sky) \\
Wavelength range & 4810--5340~\AA \\
Spectral features & H$\beta$, [OIII], Mg$b$, Fe5270 \\
Total efficiency & 14.7\% \\
Commissioned & February 1999 at WHT \\
\hline
\end{tabular}
\end{table}
\end{center}

\begin{figure}
\centerline{\psfig{figure=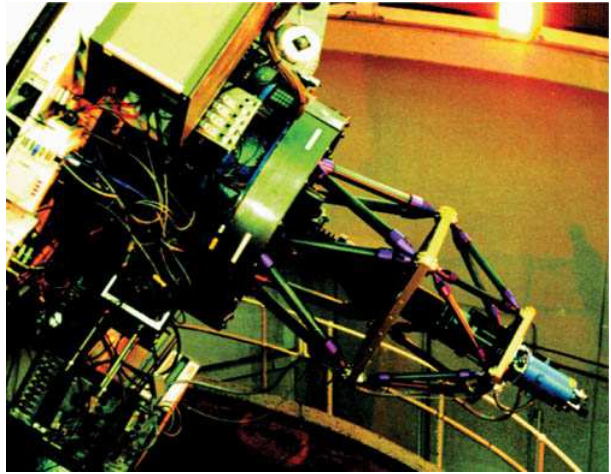,width=8cm}}
\caption{{\tt SAURON} at the Cassegrain focus of the William Herschel
Telescope.}
\end{figure}

\section{Data reduction}
The goal of the data reduction software is to produce uniformly
reduced data of high quality. Depending on its size (expressed in
terms of the effective radius $r_e$), a galaxy will be observed in (a
series of) one, two or, in some rare cases, three or more telescope
pointings. Every pointing is split in a number of individual
exposures, thirty minutes each. This implies that the reduced data-set
will consist of roughly 170,000 independent spectra, corresponding to
a large volume of raw data (of the order of 100 Gb). We are currently
in the process of building a dedicated pipeline to reduce this large
data set in an efficient way.

To obtain a large field of view, the {\tt SAURON} spectra are more
densely packed than was the case for {\tt OASIS}. This means that
special attention had to be paid to the extraction of the spectra. We
also developed tools to mosaic and merge datacubes. Furthermore,
specific software for the spatial binning of spectra in two dimensions
was developed, in order to obtain almost uniform signal-to-noise over
the field of view (Cappellari \& Copin, this conference).

\section{Sample}
We have built a complete sample of nearby E/S0/Sa of 327 galaxies,
using the following constraints:

\begin{itemize}
\vskip -\baselineskip
\vskip 15 truept
\itemsep -3pt
\item radial velocity: $cz <$ 3000 km\,s$^{-1}$
\item declination: $-6^o < \delta < 64^o$ 
\item $|b| \ge 15^o$
\item absolute magnitude: $M_B \le -18$ 
\end{itemize}
\vskip -3truept
From this, we selected a representative sample of 72 galaxies by
uniformely populating the absolute magnitude--ellipticity plane.  The
sample spans a large range of global and nuclear properties, covering
over four magnitudes in brightness $M_B$, 0.6 magnitudes in color
$(B-V)_e$, 0.35 in central Mg$_2$ index, a factor 100 in rotational
support $(V/$sigma$)^*$, a factor 10 in central cusp-slope brightness
profile and a factor 1000 in black hole mass.  Half of the
representative sample is located in `cluster' environments, while the
other half is in a less dense environment.\looseness=-2

At the time of this conference 53 of the 72 galaxies were observed
(Fig.~2). The survey should be completed after the two scheduled runs
of January and April 2002. The complete list of galaxies can be found
in de Zeeuw et al. (2002).
\begin{figure}
\centerline{\psfig{figure=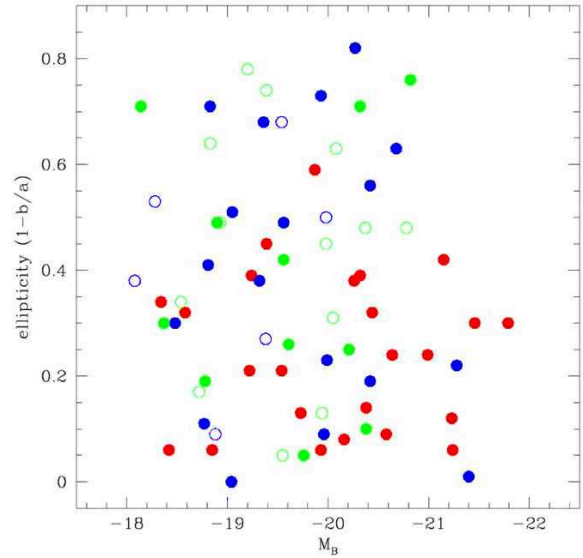,width=8cm}}
\caption{Status of the {\tt SAURON} sample in December 2001.  Observed
galaxies (filled symbols) are shown in red (E), blue (S0) and green
(Sa).  Galaxies not yet observed are represented with open symbols.}
\end{figure}

\section{Science verification}
We have performed extensive scientific verifications to check the
quality of the {\tt SAURON} measurements.  This analysis demonstrates
that the {\tt SAURON} data is of similar or even higher quality than
the best published long-slit measurements (de Zeeuw et al.\ 2002). An
example of such a comparison is shown in Fig.~3, where long-slit
kinematics along the major and minor photometric axis of the SB0
galaxy NGC3384 (Fisher 1997) is compared with similar measurements
from a simulated long-slit extracted from {\tt SAURON} maps. The RMS
deviation between the two data sets is 7 km\,s$^{-1}$ in $V$ and 8
km\,s$^{-1}$ in $\sigma$.  The important advantage of the {\tt SAURON}
measurements, as illustrated by the kinematic maps of NGC3384
(Fig.~4), is that they are two-dimensional and homogeneous\footnote{SAURON
observations of NGC3384 are presented in Bureau et al (this conference).}.

\begin{figure}
\centerline{\psfig{figure=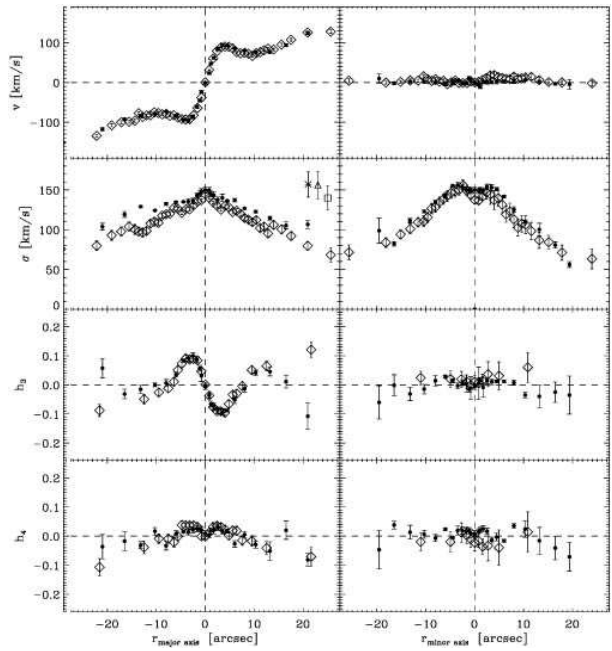,width=8cm}}
\caption{Comparison of the SAURON kinematics of the SB0 galaxy NGC3384
(dots) with long-slit measurements (open diamonds) by Fisher (1997).
The major and minor axis measurements are shown in respectively the
left and right panels.}
\end{figure}

\begin{figure}
\centerline{\psfig{figure=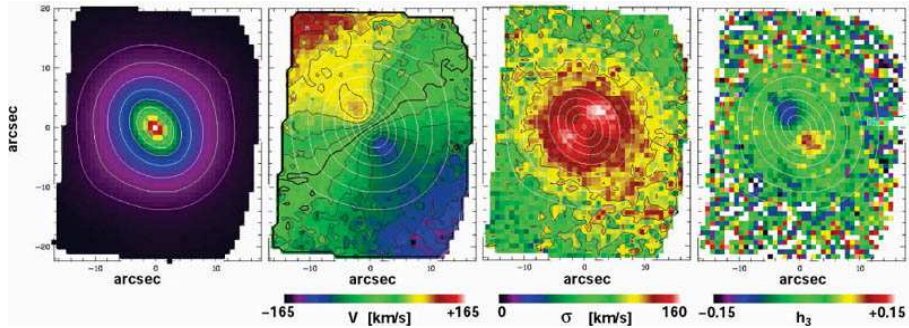,width=12cm}}
\caption{SAURON observations of NGC3384, based on a single pointing of
four 30-minutes exposures. From left to right: the reconstructed total
intensity, the mean stellar velocity, the stellar velocity dispersion
and the Gauss-Hermite moment $h_3$.}
\end{figure}

\section{Stellar kinematics}
The stellar kinematical maps (line-of-sight velocity V, velocity
dispersion $\sigma$ and the third and fourth order Gauss-Hermite
moments $h_3$, $h_4$) display a rich variety of shapes. We have
empirically classified them according to their morphology,
distinguishing the following classes (the object name between the
brackets refers to the corresponding panel in Fig.~5):
\begin{itemize}
\vskip -\baselineskip
\vskip 15 truept
\itemsep -3pt
\item {\bf Normal}. The rotation axis and the line that joins the
velocity extremes are aligned with the photometric minor and major
axes, respectively. The velocity dispersion map generally exhibits a
well-defined central peak (NGC2974).
\item {\bf Disk}. A stellar disk is clearly present in the velocity
field.  The velocity dispersion field shows a peak that is elongated
perpendicularly to the rotation axis (NGC2549).
\item {\bf Cylinder}. The velocity field displays cylindrical
rotation (NGC6501).
\item {\bf S-shaped}. The photometric minor axis and the rotation axis
are misaligned in the center (NGC1023).
\item {\bf Kinematically decoupled core}. The velocity field shows a
clearly located, well-defined and radical change in its shape
 and/or amplitude (NGC4365).
\item {\bf Strongly misaligned}. The kinematical axis is misaligned from the
photometric minor axis. An example is the bar-type rotation where the
kinematical axis in the center is aligned with the photometric
major axis in the outer regions (NGC4477).
\item {\bf Complex}. The velocity and dispersion fields display
unexpected features. This is generally due to the presence of multiple
components in the line-of-sight velocity distribution, which cannot be
taken properly into account by simple Gaussian fitting schemes.
\end{itemize}
\vskip -3truept

\begin{figure}
\centerline{\psfig{figure=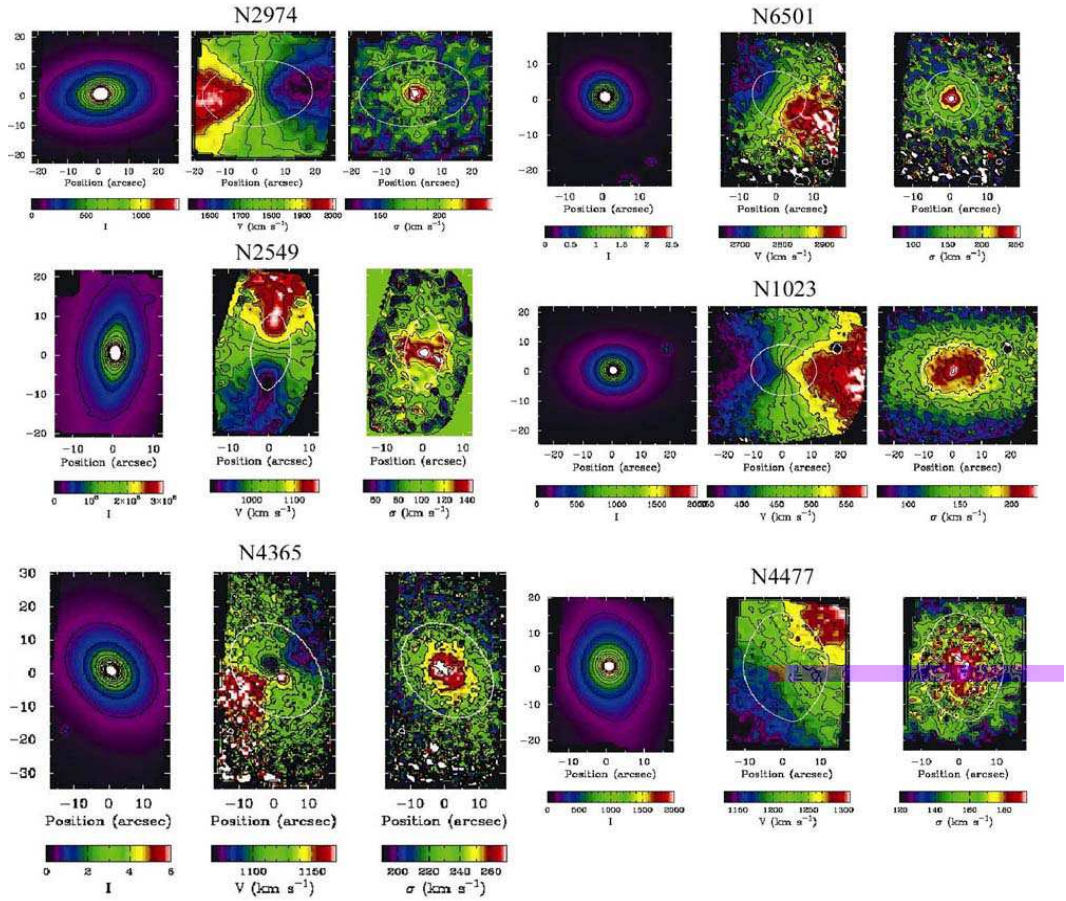,width=14cm}}
\caption{Examples of kinematical observations obtained with the {\tt
SAURON} instrument, for the galaxies NGC2974, NGC2549, NGC4365,
NGC6501, NGC1023 and NGC4477. Three panels are shown for each galaxy:
the reconstructed light distribution (left), the mean stellar velocity
field (center) and the stellar velocity dispersion field (right).  For
reference, the isophote at 0.5 $r_e$ is superimposed on the
kinematical maps; the scale is in arcsec.}
\end{figure}

We {\it quantify} these kinematical maps by means of a new tool,
`kinemetry', which extracts the relevant parameters from the velocity
field (e.g., Copin et al.\ 2001). It is similar to the photometric
ellipse fitting analysis that is often used to quantify the morphology
of light distributions in early-type galaxies.

Fully general dynamical models for some of the sample galaxies are
under construction. We use Schwarzschild's method (Schwarzschild 1979,
1982, 1993), together with a Multi-Gaussian expansion of the light
distribution (Monnet et al.\ 1992; Emsellem et al.\ 1994). The
axisymmetric three-integral software that is used (van der Marel et
al.\ 1998; Cretton et al.\ 1999) is fully general and makes no
assumptions about the distribution function or degree of anisotropy.

We applied this powerful tool first to the {\tt SAURON} observations
of M32 (Verolme et al.\ 2002). We selected M32 as a test-case because
of the availability of high-quality data, including {\tt FOS} and {\tt
STIS} observations, and reliable (dynamical) estimates of its relevant
properties (van der Marel et al.\ 1998).  Our results illustrate the
superiority of the two-dimensional {\tt SAURON}-data over multiple
position angle long-slit data: we find strong limits on the
inclination angle, which is essentially unconstrained with long-slit
data only.

Other examples of axisymmetric Schwarzschild models using {\tt SAURON}
data are given in this conference (NGC3377, Cretton et al.; NGC7332,
Falcon et al.). An extension of the Schwarzschild software to
triaxiality is nearly complete.

\section{Preliminary conclusions}

The {\tt SAURON} survey is not yet completed, and the analysis of this
large data set has only recently started. However, we can already draw
a few preliminary conclusions.  As shown in Fig.~5, stellar
kinematical maps of early-type galaxies and spiral bulges exhibit a
large variety of morphologies. While some are simple axisymmetric
systems, most are more complex than often assumed.  This confirms that
two-dimensional spectroscopy is mandatory to measure the complex
behaviour of stars and gas in these galaxies.  Furthermore, most of
these signatures do not have counterparts in the light
distribution (see Fig.~5). This implies that information contained in
the light distribution is unreliable and may even be misleading.  We
illustrate this point further in Fig.~6, where we show the velocity
fields of four E/S0 galaxies with similar light distributions. It is
clear that the kinematical behaviour is different for every object,
ranging from a fast-rotating axisymmetric object (NGC524) to a slow
and cylindrically rotating object (NGC6501), including a kinematically
decoupled core (NGC4406) and an object with non-axisymmetric rotation
(NGC4459).

\begin{figure}
\centerline{\psfig{figure=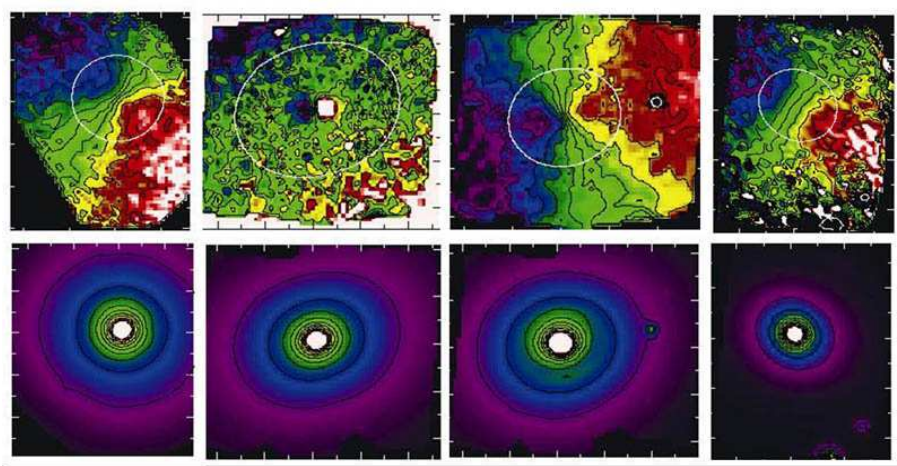,width=12cm}}
\caption{Examples of SAURON early-type galaxies that have similar 
morphology (lower panels), but very different mean stellar 
velocity field (upper panels). 
For reference, the isophote at 0.5 $r_e$ is superimposed on the kinematical 
maps; the scale is in arcsec.}
\end{figure}

A meaningful statistical analysis of the {\tt SAURON} observations
requires the complete sample and adequate tools such as
kinemetry. However, there are already indications that the kinematical
properties of nearby E/S0 galaxies do not correspond to the often
assumed simplistic two-family model, in which the giant non-rotating
triaxial ellipticals are opposed to the fast-rotating axisymmetric
faint ellipticals and S0s.  We will investigate this further by
studying the statistical incidence of kinematically decoupled cores,
stellar disks, and asymmetries, and by constructing detailed
self-consistent models for individual galaxies. This is a challenging
undertaking. In the first step towards this goal, we have successfully
built an axisymmetric three-integral Schwarzschild model for M32
(Verolme et al.\ 2002). This work shows that the presence of the {\tt
SAURON} data helps to constrain the inclination of the galaxy, which
was not possible by using long-slit observations only.  Presently, we
are developing a general triaxial Scharzschild code, which will be
applied to the many non-axisymmetric objects in the SAURON survey.

The {\tt SAURON} observations are not limited to stellar kinematics
only. They also provide line-strength and emission line maps (see
Emsellem et al., this conference). Our aim is to combine all this
information and thus gain more understanding of the current properties
of normal early-type galaxies and to derive strong constraints on their
formation and evolution. As an example, our study of the line-strength
maps of NGC4365 show that its kinematical decoupled core (Fig.~1) is a
long-lived structure that has almost the same age as the rest of the
galaxy (Davies et al., 2001).


\begin{references}
\reference Bacon R., Adam G., Baranne A., Courtes G., Dubet D., Dubois
           J.P., Emsellem E., Ferruit P., Georgelin Y., Monnet G.,
           Pecontal E., Rousset A., Say F., 1995, A\&AS, 113, 347
\reference Bacon R., Emsellem E., Copin Y., Monnet G., 2000, in {\sl Imaging 
           the Universe in Three Dimensions}, eds W.\ van Breugel \& J.\ 
           Bland--Hawthorn, ASP Conf.\ Ser.\ 195, 173
\reference Bacon R., Copin Y., Monnet G., et al. 2001, MNRAS, 326, 23
\reference Bacon R., Emsellem E., Combes F., Copin Y., Monnet G., Martin P.,
           2001, A\&A, 371, 409
\reference Bureau M. et al, this conference
\reference Cappellari M., Copin Y., this conference
\reference Copin Y., Bacon R., Bureau M., Davies R.L., Emsellem E., Kuntschner 
           H., Miller B.W., Peletier R.F., Verolme E.K., de Zeeuw P.T.,
           2001, SF2A 2001, eds F.\ Combes, D.\ Barret, F.\ Th\'evenin,
           EDPS Conf.\ Series in Astronomy \& Astrophysics, 289--292
\reference Cretton N., de Zeeuw P.T., van der Marel R.P., Rix H.-W.,         
           1999, ApJS, 124, 383 
\reference Cretton N. et al., this conference 
\reference Davies R., Kuntschner H., Emsellem E., Bacon R., Bureau M., Carollo
           C.M., Copin Y., Miller B.W., Monnet G., Peletier R.F., Verolme E.K.,
           de Zeeuw P.T., 2001, ApJL, 548, 33
\reference Emsellem E., Monnet G., Bacon R., 1994, A\&A, 285, 739
\reference Emsellem E. et al., this conference 
\reference Falcon J. et al., this conference 
\reference Fisher D., 1997, AJ, 113, 950
\reference van der Marel R.P., Cretton N., de Zeeuw P.T. \& Rix H.-W., 
              1998, ApJ, 493, 613 
\reference Monnet G., Bacon R., Emsellem E., 1992, A\&A, 253, 366
\reference Schwarzschild M., 1979,, ApJ, 232, 236
\reference Schwarzschild M., 1982, ApJ, 263, 599
\reference Schwarzschild M., 1993, ApJ, 409, 563
\reference Verolme E.K., Cappellari M., Copin Y., van der Marel R.P., Bacon R.,
           Bureau M., Davies R.L., Miller B.M., de Zeeuw P.T., 2002, MNRAS, 
           submitted (/astro-ph/0201086)
\reference de Zeeuw P.T., Bureau M., Emsellem E., Bacon R., Carollo C.M., Copin
           Y., Davies R.L., Kuntschner H., Miller B.W., Monnet G., Peletier 
           R.F., Verolme E.K., 2002, MNRAS 329, 513
\end{references}
\end{document}